# AI Blob! LLM-Driven Recontextualization of Italian Television Archives

Roberto Balestri

roberto.balestri2@unibo.it

0009-0000-5008-2911

Dipartimento delle Arti, Università di Bologna

## Abstract

This paper introduces *AI Blob!*, an experimental system designed to explore the potential of semantic cataloging and Large Language Models (LLMs) for the retrieval and recontextualization of archival television footage. Drawing methodological inspiration from Italian television programs such as Blob (RAI Tre, 1989–), *AI Blob!* integrates automatic speech recognition (ASR), semantic embeddings, and retrieval-augmented generation (RAG) to organize and reinterpret archival content. The system processes a curated dataset of 1,547 Italian television videos by transcribing audio, segmenting it into sentence-level units, and embedding these segments into a vector database for semantic querying. Upon user input of a thematic prompt, the LLM generates a range of linguistically and conceptually related queries, guiding the retrieval and recombination of audiovisual fragments. These fragments are algorithmically selected and structured into narrative sequences producing montages that emulate editorial practices of ironic juxtaposition and thematic coherence. By foregrounding dynamic, content-aware retrieval over static metadata schemas, *AI Blob!* demonstrates how semantic technologies can facilitate new approaches to archival engagement, enabling novel forms of automated narrative construction and cultural analysis. The project contributes to ongoing debates in media historiography and AI-driven archival research, offering both a conceptual framework and a publicly available dataset to support further interdisciplinary experimentation.

# 1. Introduction

In recent years, archival methodologies have been undergoing a deep transformation, driven by advancements in artificial intelligence and machine learning. Traditional metadata-based cataloging and retrieval techniques are increasingly being supplemented—or in some cases, entirely redefined—through AI-powered automation and semantic search technologies (Colavizza *et al.* 2021). Although these innovations have significantly improved the searchability of archival materials by focusing on content indexing, emerging discussions in the field, as seen in Allan et al. (2024), indicate a growing interest in approaches that also support creative reuse.

This paper presents *AI Blob!*, an experimental system that uses the capabilities of Large Language Models (LLMs) and Retrieval-Augmented Generation (RAG) in television archives. Drawing inspiration from the montage-based editorial techniques of the iconic Italian television program *Blob* (RAI Tre, 1989-), *AI Blob!* explores how semantic cataloging, AI-assisted retrieval and LLMs generative capabilities can help in recontextualizing archival footage.

The dataset produced and utilized for this project consists of 1,547 automatically transcribed videos, from which 212,696 distinct sentences were extracted and embedded using a multilingual multimodal embedding model. This structure allows for future expansion, including the integration of image embeddings, which would enable a richer and more comprehensive cataloging system, enhancing both semantic retrieval and cross-modal search capabilities. We are making this dataset, along with the transcriptions and the pre-filled vector store, publicly available to encourage further experimentation and development in the field of AI-assisted media historiography.[1] The project's source code is likewise openly available.[2] Additionally, a selection of videos generated by *AI Blob!* can be watched in the dedicated YouTube playlist.[3]

---

[1] https://zenodo.org/records/15071951

## 2. Related Works

The concept of embedding has been fundamental in advancing artificial intelligence applications, particularly in natural language processing (NLP) and information retrieval. Embeddings map textual or multimedia content into high-dimensional vector spaces, allowing models to capture semantic relationships beyond exact keyword matches. Early embedding models such as Word2Vec (Mikolov *et al*. 2013) and GloVe (Pennington *et al.* 2014) laid the foundation for semantic search by representing words based on their contextual usage. More recent advancements, including transformer-based embeddings (Devlin *et al.* 2019, Radford *et al.* 2018) and cross-lingual representations (Conneau *et al.* 2020), have significantly improved AI's ability to understand and retrieve information from large-scale textual and audiovisual datasets. In archival research, embeddings facilitate efficient semantic retrieval, allowing AI systems to recognize thematic and conceptual similarities between media fragments, even when the phrasing or visual representation varies.

LLMs rely on embeddings to represent words and sentences as vectors that capture semantic relationships, enabling advanced language understanding and generation (Vaswani *et al*. 2017, Brown *et al*. 2020, OpenAI *et al*. 2023). These models have demonstrated significant potential, offering new possibilities for narrative analysis (Balestri and Pescatore 2025), automated storytelling (Zhao et al. 2023) and creative reinterpretation of filmic material (Balestri *et al.* 2024).

---

[2] https://github.com/robertobalestri/AI-Blob-MM16

[3] https://www.youtube.com/playlist?list=PLF3Gpk1ZBOk2Y334h8WcCAi4VBopu4lXf

In recent years, artificial intelligence has reshaped archival methodologies, with projects like *AI4Media*,[4] demonstrating the potential of speech recognition, NLP, and computer vision to improve audiovisual archive accessibility (Bazán-Gil 2023). A 2022 study by Radiotelevisione Italiana (RAI), conducted as part of the *AI4Media* project, found that while AI applications in the media industry are no longer an emerging trend, they have yet to become a widely adopted industrial standard (Bruccoleri *et al*. 2022).

AI-powered tools have enhanced metadata tagging, content indexing, and retrieval, as seen in RAI *Teche*'s AI-driven metadata extraction and INA's use of facial and speech recognition. However, most of these approaches focus on retrieval and structuring, rather than reinterpretation of archival material (Bazán-Gil 2023). An early effort toward the automatic recontextualization of television archives was undertaken by the British Broadcasting Corporation (BBC) with the program *BBC Made by Machine: When AI Met the Archive* (BBC 2018).[5] In this project, clips from BBC programs were cataloged by AI and automatically edited based on semantic similarity. However, due to the absence of LLMs' advanced semantic understanding, the system relied solely on surface-level similarity, resulting in a montage that juxtaposed clips without deeper contextual comprehension.

Italian TV archives, among the richest in Europe, have long been repurposed for thematic compilations and commercial reuse (Barra and Scaglioni 2012). Programs like *Blob* (RAI Tre 1989–) or *Da Da Da* (Rai Uno 2010-2020) pioneered the use of montage editing to satirically reinterpret media, exposing contradictions through the juxtaposition of archival clips.

*AI Blob!* expands on traditional archival methods by using LLMs to enable the dynamic recontextualization of television footage. Moving beyond static metadata and keyword searches, it employs semantic embeddings and Retrieval-Augmented Generation (RAG) (Lewis *et al.* 2020) to construct narrative sequences that reflect thematic links and emulate editorial choices in television montage.

---

[4] https://www.ai4media.eu/ (visited: 20/03/2025)

[5] https://www.bbc.co.uk/programmes/b0bhwk3p (visited: 20/03/2025)

## 3. Dataset Preparation

The dataset for *AI Blob!* was curated from two primary sources: the *ITTV dataset*[6] (Mezza *et al.* 2023) and *Indimenticabile TV*,[7] a YouTube channel featuring excerpts from classic Italian television.

*ITTV Dataset*, originally developed for automatic TV genre classification, includes 2,625 manually annotated YouTube videos across seven categories: Cartoons, Commercials, Football, Music, News, Talk Shows, and Weather Forecast. From *Indimenticabile TV* channel we were able to download 461 different videos in January 2025.

To ensure consistency, two key filtering steps were applied:

- Any videos from the ITTV dataset that were no longer accessible on YouTube were removed.
- Videos with a majority of non-Italian content were identified using FastText, a deep learning-based language classification model (Joulin *et al.* 2016a; Joulin *et al.* 2016b)[8] and removed.

After the filtering process, we ended up with a total of 1,547 unique videos.

---

[6] https://zenodo.org/records/8027327 (visited: 20/03/2025)

[7] https://www.youtube.com/@indimenticabiletv (visited: 20/03/2025)

[8] https://fasttext.cc/docs/en/language-identification.html (visited: 20/03/2025)

## 4. Transcription and Embedding

For each video, Automatic Speech Recognition (ASR) was performed using WhisperX[9] (Bain *et al*. 2023), selected for its efficiency, precise word timestamping, and ability to align transcriptions with audiovisual content. Once transcribed, the text was processed to extract 212,696 distinct sentences, ensuring clean and meaningful segmentation for later embedding. To achieve this, two deep learning models were employed: Punkt Sentence Tokenizer (Bird *et al.* 2009: 112-113), a rule-based tokenizer from the Natural Language Toolkit (NLTK),[10] and xlm-roberta_punctuation_fullstop_truecase,[11] , a transformer-based model designed to restore punctuation and proper casing, enhancing the coherence of extracted sentences.

These processed sentences were then embedded using Embed Multilingual V3,[12] a multimodal embedding model developed by Cohere. The embeddings were stored in ChromaDB, a specialized vector database optimized for fast semantic search and retrieval.

## 5. Methodology: Script Production and Narrative Construction

Inspired by the Italian TV program *Blob* (RAI Tre, 1989–), which takes "a caustic and satiric perspective to both television itself and society" and often creates thematic episodes around specific topics (Barra and Scaglioni 2012), the approach combines semantic retrieval, generative AI, and interpretative logic to automatically select, filter, and reorganize archival materials into structured

---

[9] https://github.com/m-bain/whisperX (visited: 20/03/2025)

[10] https://github.com/nltk/nltk/ (visited: 20/03/2025)

[11] https://huggingface.co/1-800-BAD-CODE/xlm-roberta_punctuation_fullstop_truecase (visited: 20/03/2025)

[12] https://huggingface.co/Cohere/Cohere-embed-multilingual-v3.0 (visited: 20/03/2025)

narratives. Beginning with the user's selection of a central theme, namely the title of the episode, the methodology follows sequential stages as explained in the next paragraphs (see Figure 1 for a visual reference).

## 5.1 Thematic Exploration and Query Generation

The first stage involves generating thematic ideas that introduce ironic, absurd, or paradoxical perspectives on the selected central theme. Following the principles of lateral thinking (De Bono and Zimbalist 1970), which emphasize creative exploration beyond traditional or linear reasoning, this stage deliberately seeks unconventional viewpoints to provoke novel and unexpected interpretations. From these thematic ideas, the system generates a set of related phrases to be used as queries for exploring the semantic database. This ensures diverse entry points for semantic retrieval, broadening the scope of archival material available for satirical reinterpretation.

## 5.2 Semantic Retrieval from Vector Database

The queries generated are used to retrieve relevant sentences from the vector database. The retrieval process employs logic to exclude sentences previously selected by other queries, ensuring diverse and non-redundant content selection.

## 5.3 Evaluation and Filtering for Ironic or Thematic Relevance

The retrieved sentences are computationally evaluated for their potential to produce ironic or satirical effects when removed from their original contexts. This evaluation process is conducted by the LLM, which assigns each sentence an irony score on a scale from 1 to 10. The scoring is based on multiple interpretive dimensions, including semantic ambiguity, humorous dissonance when isolated from context or paradoxical formulations.

In addition to irony, the system also assesses each sentence's relevance to the thematic focus of the project, using a parallel scoring system (also on a 1–10 scale). The relevance score captures how conceptually or topically aligned a sentence is with the overarching theme, accounting for latent or oblique connections that may enhance the coherence or critical framing of the final narrative.

The system retains all sentences that meet either of the two criteria: a high irony score or a high thematic relevance score (above predefined thresholds).

## 5.4 Algorithmic Narrative Segmentation of Filtered Sentences

After the initial filtering phase, retained sentences undergo an algorithmically driven narrative segmentation. This computational procedure categorizes sentences directly based on their measured ironic and thematic relevance scores. The segmentation process leverages the statistical distribution of these scores, dynamically grouping sentences into narrative sections corresponding to distinct montage functions. Specifically, the computational procedure identifies appropriate thresholds through percentile-based segmentation:

- Introduction: Comprises sentences characterized by relatively lower irony yet higher thematic relevance, facilitating a gentle but clearly ironic thematic entry.
- Build-up: Organized progressively, these segments escalate the irony score, gradually intensifying ironic tension. They serve to systematically develop the thematic and ironic depth of the montage, laying groundwork for the narrative climax.
- Climax: Consists exclusively of sentences with the highest ironic scores, maximizing narrative intensity through pronounced absurdity, paradox, or extreme contradiction.
- Conclusion: Features sentences exhibiting moderated irony and balanced thematic relevance, thus providing narrative closure, reflective irony, or summarizing commentary.

## 5.6 Computational Ordering and Strategic Juxtaposition within Narrative Segments

Following segmentation, each narrative section is ordered by the LLM. Sentences within each segment are sequenced to deliberately maximize ironic and semantic contrasts, using the predetermined narrative purpose of each segment (introduction, progressive tension, climax, and conclusion) as explicit guidelines. Computational ordering ensures that sentences identified as serious or thematically coherent are juxtaposed strategically against absurd, contradictory, or ironic statements, thereby intensifying narrative tension and enhancing comedic or satirical effect within each structured segment of the montage.

## 5.7 Final Video Assembly and Audiovisual Editing

The concluding stage involves assembling the ordered narrative segments into a unified audiovisual montage. Selected archival clips are retrieved from original sources based on transcription timestamps. To ensure seamless transitions, audio fade-in and fade-out effects are applied to each segment. An introductory video sequence is added at the beginning to provide thematic context. Finally, all elements are concatenated into a cohesive video. Audio normalization and dynamic range compression are performed during assembly to ensure consistent audio quality.

# 6. Limitations, Future Work and Conclusion

## 6.1 Limitations

Despite its promising results, *AI Blob!* presents several limitations. First, word alignment in the automatic speech recognition process can be imprecise, leading to occasional mismatches between spoken text and video content. Second, while LLMs guide phrase selection, the extracted clips don't always achieve thematic coherence or effective ironic contrast. Third, the system analyzes content unimodally, overlooking visual elements crucial to television's multimodal language. This hinders the recreation of *Blob*'s signature style, which often relied on ironic audio-visual juxtapositions (Barra and Scaglioni 2012). Finally, the dataset's relatively small size (1,547 videos) limits thematic specificity and narrows the scope for in-depth satirical montage construction.

## 6.2 Future Work

Two major directions could enhance the system. First, incorporating multimodal embeddings would allow better integration of visual content, capturing visual irony and improving montage cohesion. Second, expanding the dataset would increase thematic diversity and enable more granular narrative construction. Additional developments may involve moving from the current approach of filtering a collection of clips to a more sequential method that builds the narrative progressively by selecting each new sentence based on its relationship to previously selected content during the vector store retrieval phase.

## 6.3 Conclusion

*AI Blob!* represents a step forward in using AI for audiovisual archival research and creative reuse. It goes beyond metadata, employing semantic search, language models, and automated structuring to support meaning-driven montage, echoing the critical and satirical ethos of *Blob*. The release of the dataset, transcriptions, and vector store promotes further experimentation in AI-driven media historiography. While not (yet) a substitute for human editorial insight, *AI Blob!* opens new avenues for collaborative, critical reinterpretation of archival content—offering a model for computational creativity that respects television's legacy while embracing AI's potential.

# Bibliografia